%% file: cosmopub.tex
\documentclass[12pt]{article}
\usepackage{graphics}
\input epsf
\input defs

\newcommand{\be}{\begin{eqnarray}}
\newcommand{\ben}{\begin{eqnarray}\nonumber}
\newcommand{\ee}{\end{eqnarray}}

\begin{document}

\title{Einstein's Equations and a Cosmology with Finite Matter}
\author{
L. Clavelli\footnote{lclavell@bama.ua.edu}$\;$\footnote{emeritus at
Dept. of Physics and Astronomy, Univ. of Alabama, Tuscaloosa AL 35487} and
Gary R. Goldstein\footnote{gary.goldstein@tufts.edu}\\
Dept. of Physics and Astronomy, Tufts University, Medford MA 02155}

\date{Mar 4, 2015}
\maketitle

\begin{abstract}
We discuss various space-time metrics which are compatible with Einstein's equations and a previously suggested cosmology with a finite total mass \cite{ClavelliGoldstein}. In this alternative cosmology the matter density was postulated to be a spatial delta function at the time of the big bang thereafter diffusing outward with constant total mass.
This proposal explores a departure from standard assumptions that the big bang occured everywhere at once or was just one of an infinite number of previous and later transitions. 
\end{abstract}

keywords: Cosmology, Solutions of Einstein's equations, Finite Mass Universe, preferred frame

DOI: 10.1142/SO217751X15500682

\section{Introduction}

It is currently widely believed that the universe is, on large scales, isotropic and homogeneous although there are many nagging discrepancies from this standard cosmological model including many unexplained asymmetries and 
correlations \cite{Perivolaropoulos}.
In this paper we address four aspects of the standard cosmological model deserving of discussion.  

\begin{enumerate}

\item{In the standard cosmological model and in any homogeneous cosmology with a non-zero 
probability per unit space-time volume to produce an individual of any species, there results an 
infinite replication of each possible human, quasi-human, and monster individual.}  We referred to this bizarre property as infinite cloning. 

\item{The standard cosmological model is vague with regard to initial conditions.}
Although well-founded studies \cite{Ateam} have deduced that matter must have originated at a finite time in the past, many authors continue to seek a cosmology extending into the infinite past. 
\item {In the standard cosmological model, the matter density is homogeneous on large scales.}
If matter is homogeneous on large scales there should be no larger structures.  Surprisingly, however, clusters of galaxies have been found with dimensions as great as any probed scale.
\item The standard cosmological model is often described in the inflationary era as an infinite de-Sitter space with a large cosmological constant.  The current era is then described as another infinite de-Sitter space with a much smaller cosmological constant.  Relativity demands that the transition cannot occur everywhere at once but should have an expanding bubble topology which is not evident in the standard model.
\end{enumerate}

 The first of these four issues may be primarily philosophical in nature as long as the infinite cloning is at causally disconnected patches of space-time.  Nevertheless it is, perhaps, interesting to ask whether there is a viable cosmological model without this cloning property.  This was the motivation of ref \cite{ClavelliGoldstein}. In addition, this infinite replication of everything is also at the 
heart of the measure problem which is acknowledged to be a serious puzzle in standard cosmology. 

It has been suggested that the question of an origin of time is also somewhat philosophical since, even if time has an origin, it may be at such a great time in the past as to be for all practical purposes infinitely remote.  To this one could answer that a model that addresses the initial condition problem is potentially better than one that avoids the question.   In quantum physics the initial state can be ``prepared" in any quasi-stationary state after which the evolution of the system follows from the equations of motion and quantum jumps to the ground state or other quasi-stationary states are allowed. 
An initial state in a slow roll between two quasi-stationary states would be hard to reconcile with quantum theory.

Thirdly, even if the homogeneity of the matter distribution on large scales becomes consistent with observation, one should ask whether there are viable models where inhomogeneity sets in at larger scales such as the model we propose.  Faced with an apparent symmetry of matter, it is a time-honored tradition in physics to construct a parameter-dependent model that agrees with the apparent symmetry only in some limit thus replacing a theoretical question with the experimental question of observational constraints.

Much effort has been expended to resolve or soften the big bang singularity while little 
attention has been given to dealing with the possibility of an actual singularity. Taking the 
time of the big bang at $t=0$, the proposal of ref.\cite{ClavelliGoldstein} is that the matter 
density $\rho_m(\vec{r},t)$ satisfies: 
\be 
\lim_{t \rightarrow 0} 
\,\rho_m(\vec{r},t) = \left\{ \begin{array}{ll} \infty &\quad r=0\\ 0 & \quad r\ne 0 \end{array} 
\right. 
\label{rhom1} 
\ee 
with, at all positive times, 
\be
    \int d^3 r \,\rho_m(\vec{r},t) = M  \quad .
\label{rhom2}
\ee
These equations define $\rho_m{(\vec{r},t)}$ as proportional to a spatial delta function at time $t=0$:
\be 
    \lim_{t \rightarrow 0}\, \rho_m(\vec{r},t) = M \delta^3(\vec{r}) \quad .
\ee
A particular example of a matter density satisfying these equations is
\be 
    \rho_m (r,t) = \frac{M}{(R {\sqrt\pi}\,a(t))^3} e^{-r^2/(Ra(t))^2} \quad .
\label{rhom}
\ee
where
the scale factor $a(t)$ is a function of time that vanishes at $t=0$.  The model becomes consistent with standard cosmological model homogeneity at distances such that $r/a(t)$ is much smaller than the parameter $R$.  Thus $R$ plays the role of a scale of inhomogeneity.

 In the co-moving frame the matter density takes the time independent form
\be 
    \rho_c (r) = \frac{M}{(R {\sqrt\pi})^3} e^{-r^2/R^2} \quad .
\label{rhoc}
\ee

A Gaussian density is the ground state of the three dimensional harmonic oscillator so the model could have the effect of matter confined by a quadratic potential. 
 The Gaussian fluctuations of the free harmonic oscillator together with the assumed flat primordial spectrum is key to the standard cosmological prediction of the acoustic peaks in the cosmic background radiation (CBR)
\cite{HarrisonZeldovich}. However, at present, we limit our attention to 
purely classical considerations as in the Friedmann-Robertson-Walker (FRW) metric. 
 A possible future extension of the current treatment might be to take $\rho_m$ to be a temperature dependent superposition of
excited states which could bring in the angular variations suggested in ref.\,\cite{ClavelliGoldstein}. The number density of equivalent nucleons is $1/m_N$ of this matter density (including dark matter) and the number density of photons is proportional to the number density of nucleons.   
The photon energy density proportional in our model to eq.\,\ref{rhom} with an extra inverse factor of $a(t)$ also satisfies eq.\,\ref{rhom1} although the total energy in photons is time dependent due to red-shift. 

As we shall show, the model predicts a flat space-time at very small (and very large) $r/R$.
In regions of flat space, the equation of continuity for $\rho_m$
\be
     \vec{\nabla} \cdot (\rho_m \vec{v}(r)) + \frac{\partial \rho_m }{\partial t} = 0
\label{eqofcontinuity}
\ee
implies, independent of $R$, Hubble's law:
\be
     \vec{v}(r) = \vec{r}\; \frac{\dot{a}}{a} \quad .
\label{HubbleFlow}
\ee
The decceleration parameter is
\be
    q = - \frac{\ddot{a} a}{\dot{a}^2} \quad .
\ee

\begin{figure}[htbp]
\begin{center}
\epsfxsize= 3in 
\leavevmode
\epsfbox{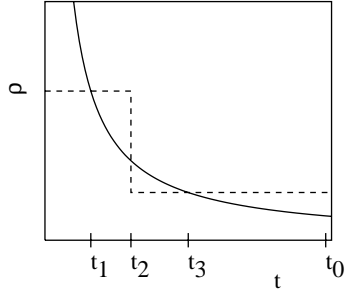}
\end{center}
\caption{Energy densities of matter plus radiation at peak (solid curve) and dark energy (discontinous dashed curve) 
are schematically represented as a function of time.  Relative times and energies are not drawn to scale.}
\label{energies}
\end{figure}

In the matter dominated regime of a Friedmann-Robertson-Walker model, the scale factor should vary as $t^p$ with $p=2/3$ or, in the relativistic regime $p=1/2$.  
In the vacuum energy dominated regime of that model, the scale factor should vary as $e^{H_v t}$ with $H_v$ being the vacuum Hubble parameter.  

A parametrization of an early universe scale factor vanishing at $t=0$ is
\be
    a(t)= b\,(\Omega_{DE} (e^{H_{v1} t} - 1) +\Omega_r (H_{r1} t)^{1/2}+\Omega_m (H_{m1} t)^{2/3}) 
\label{scalefac1}
\ee
with time dependent dark energy, radiation, and matter fractions given by $\Omega_i$.  
After a transition to mild inflation at some time $t_2$, the scale factor should rise to unity as time approaches the present time $t_0=13.8\,{\displaystyle Gyr}$.
\be
     a(t)= \frac{\Omega_{DE} (e^{H_{v2} t} - 1) +\Omega_r (H_{r2} t)^{1/2}+\Omega_m (H_{m2} t)^{2/3}}{\Omega_{DE} (e^{H_{v2} t_0} - 1)+\Omega_r (H_{r2} t_0)^{1/2} +\Omega_m (H_{m2} t_0)^{2/3}} \quad .
\label{scalefac2}
\ee
The constant $b$ can be chosen to make the scale factor continuous through the transition.
\be\nonumber
    b &=& \frac{1}{\Omega_{DE} (e^{H_{v1} t_2} - 1) +\Omega_r (H_{r1} t_2)^{1/2}+\Omega_m (H_{m1} t_2)^{2/3}}\\
    &\cdot&\,\frac{\Omega_{DE}(e^{H_{v2} t_2} - 1) +\Omega_r (H_{r2} t_2)^{1/2}+\Omega_m (H_{m2} t_2)^{2/3}}{\Omega_{DE} (e^{H_{v2} t_0} - 1) +\Omega_r (H_{r2} t_0)^{1/2}+ \Omega_m (H_{m2} t_0)^{2/3}}
\ee

This simplified parametrization ignores some fine points such as the persistence of relativistic matter for some time beyond the transition to mild inflation.  

The current model has eight free parameters
($M, R, H_{v1}, H_{r1},H_{m1}, H_{v2},H_{r2}, H_{m2}$) the last six of which are also free parameters in the standard cosmological model along with its fine-tuned slow roll potential.
 
With these choices, matter is born in the big bang with an outward velocity and inward acceleration.  

Initially there is a large vacuum energy density $\Lambda_1= (2 \cdot 10^{16} {\displaystyle GeV})^4$.
This estimate of $\Lambda_1$ is predicated on the solution of the monopole problem which requires that the GUT scale transition happen during or before the era of rapid inflation.  

The subsequent history, sketched in fig.\ref{energies}, is marked by the following critical times:

\begin{enumerate}
\item At time $t_1$ the initially positive deceleration parameter, $q$, vanishes becoming afterwards negative.
\item At time $t_2$ there is a phase transition to the current mild vacuum energy
$\Lambda_2 \approx (2\cdot 10^{-12} {\displaystyle GeV})^4 \quad$. At this point, the matter density (including radiation) again dominates over the vacuum energy.  In multiverse theory this would be the end point of a probably long sequence of intermediate vacuum energies. The phenomenological advantages of inflation (flatness etc.) require that
$a(t_2)/a(t_1) \approx e^{60}$. 

\item Above some time $t_3$ and up to the present the mild vacuum energy again dominates over the matter density.

\item If the metric is to avoid space-time inversion in which the radial and time components of the metric tensor change sign, there must be a future transition at time $t_4$ to zero vacuum energy and a constraint on the total mass of the universe in our inhomogeneous model.
\end{enumerate}

The main question of this
paper is whether the assumptions of eqs.\ref{rhom1} are consistent with Einstein's equations or whether they would require modifications of Einstein's theory.
 
According to Einstein's theory, there is a proportionality between the energy-momentum tensor, $T_{\mu \nu}$ and the Einstein tensor, $G_{\mu \nu}$, which is itself a function of the metric tensor $g$.
\be
     G^{\mu} _{\nu}(g) = - 8 \pi G_N T^{\mu} _{\nu}  \quad .
\ee
By a choice of units we suppress the factor of $8 \pi G_N$. The energy density 
is then 
\be
     \rho(\vec{r},t)=T^{4}_{4} = -G^{4}_{4} \quad .
\ee

In general, the equations have a solution for any postulated energy-momentum tensor.  Rather than attack this problem head-on with a supercomputer, we prefer to examine simple choices for a metric which lead analytically to a matter density satisfying eqs.\ref{rhom1},\ref{rhom2}.  We use standard packages in Mathematica and Maple to perform this calculation.
  
\section{A modified continuous Schwarzschild-deSitter metric}

We consider a metric of the form
\be
    ds^2 = -g_{44}^{-1}(r,t) dr^2 + r^2 d\Omega^2 + g_{44}(r,t) dt^2 \quad .
\label{metric}
\ee
With this metric the non-zero components of the Einstein tensor are
\be\nonumber
   G^{4}_{4} &=& -(1/r^2)(1 + g_{44}(r,t) + r g_{44}^\prime(r,t))\\\nonumber
   G^{r}_{r} &=& G^{4}_{4}\\\nonumber
   G^{\theta}_{\theta} &=& \frac{{\dot{g}_{44}(r,t)}^2}{g_{44}(r,t)^3} - \frac{{\ddot{g}_{44}(r,t)}}{2 {g_{44}(r,t)}^2}-\frac{g_{44}^\prime (r,t)}{r}
       -\frac{g_{44}^{\prime \prime}(r,t)}{2}\\\nonumber
   G^{\phi}_{\phi} &=& G^{\theta}_{\theta}\\\nonumber
   G^{r}_{4} &=& {\dot {g}_{44}(r,t)}/r\\
   G^{4}_{r} &=& -\frac{{\dot {g}_{44}(r,t)}}{r g_{44}(r,t)^2} \quad .
\label{EinsteinTensor}
\ee
Here, prime refers to a derivative with respect to the radial coordinate and dot refers to a derivative with respect to time.

The proposed metric and its resulting Einstein tensor share some properties with the Vaidya metrics
\cite{Vaidya} which are defined as
\be
    ds^2_{\displaystyle{Vaidya}} = -(1-2h(u)/r)(du)^2 + r^2 d\Omega - 2 du dr
\ee
where $u + 2 h \, ln\,{(\frac{r}{2 h} -1)} = t -r \quad .$

As in our eq.\,\ref{EinsteinTensor} the Vaidya metrics result in off-diagonal Einstein Tensor elements (mass currents) and 
lead to an equality of the $G^{4}_{4}$ and $G^{r}_{r}$, a property of ``null dusts". The Vaidya metrics were invented to model a black hole accreting or radiating. They represent examples of inhomogeneous, time dependent matter densities which, presumably however, cannot be made to satisfy our eqs.\,\ref{rhom1},\ref{rhom2}. 

The {\bf Schwarzschild solution} for a black hole of mass m corresponds to
\be
    g_{44}(r,t) = -1 + 2 m/r \quad .
\label{Schwarzschild}
\ee
According to Birkhoff's theorem this time-independent metric is the unique zero vacuum energy solution of Einstein's equations in the region external to a mass $m$.  If the mass is point like, the solution corresponds to a vanishing energy-momentum tensor at positive $r$ but to a delta function singularity comprising mass $m$ at the origin.

The {\bf Schwarzschild-deSitter} solution generalizes this to a space with vacuum energy density $\Lambda$:
\be
    g_{44}(r,t) = -1 + 2 m/r + r^2 \Lambda/3 \quad .
\label{SdSmetric}
\ee
It corresponds again to vanishing density and pressure of matter in the external region but to a constant vacuum energy density and pressure.  { Both of these metrics define a light-trapping region where $g_{44}(r,t)$ changes sign.  They require either restricting the region of applicability or attaching a complementary patch of space time.}  
We will see a similar phenomenon in the metrics we study.  

In this section we further generalize the Schwarzschild-deSitter metric by replacing the point-like mass $m$ with
a function of space and time.
\be
      g_{44}(r,t) = -1 + h(r,t) + r^2 \Lambda/3 \quad .
\label{metric2}
\ee

If $h(r,t)$ extends continuously over all space there is no contradiction with Birkhoff's theorem.
We aim to satisfy Einstein's equations with a matter density similar to that of eq.\,\ref{rhom}. To this end we might consider
\be
     h(r,t) = c_0 \frac{r^2}{(R a(t))^3} e^{-(r/(R a(t)))^2} \quad ,
\label{aitch}
\ee
according to which the energy density from eq.\,\ref{EinsteinTensor} is
\be
     \rho(r,t)=-G^{4}_{4}= \Lambda + \frac{c_0}{(R a(t))^3}(3-2 (r/(R a(t)))^2 e^{-(r/(R a(t)))^2} \quad .
\label{rho3}
\ee     
In this case the matter density proportional to $c_0$ becomes negative at large $r$ and integrates to zero.

In the following section we study the possibility of avoiding this by multiplying $h(r,t)$ by an
infinite series in $(r/(Ra(t))^2$.  In section 4 we consider restricting $h(r,t)$ to the region of
positive $\rho$ in eq.\,\ref{rho3}.
The matter density would then describe an expanding bubble.

\section{A positive definite density with infinite range}

In this section we avoid the appearance of negative energy densities by writing the infinite series
\be
    h(r,t)=\frac{r^2}{(R a(t))^3} e^{-(r/(R a(t)))^2} \sum_{n=0}^{\infty} c_n (\frac{r}{R a(t)})^{2n} \quad .
\ee
Then,
\be
    \dot{g}_{44}(r,t) = \frac{\dot{a}(t)}{a(t)} \frac{r^2}{(R a(t))^3} e^{-(r/(R a(t))^2} \sum_{n=0}^{\infty}
                 c_n (\frac{r}{R a(t)})^{2n}(-(2 n + 3)+2 (\frac{r}{R a(t)})^2)  \quad .
\ee   
We can choose
\be
     c_n = c_0 \frac{\Gamma(5/2)}{\Gamma(5/2+n)} 
\ee
so that
\be
    \frac{1+g_{44}(r,t)+r g_{44}^\prime (r,t)}{r^2} = \Lambda + \frac{3 c_0}{(R a(t))^3} e^{-(r/(R a(t))^2}
\ee
and
\be
        \dot{g}_{44}(r,t) = -3 c_0 \frac{\dot{a}(t)}{a(t)} \frac{r^2}{(R a(t))^3} e^{-(r/(R a(t))^2}
   \quad .
\ee
One can see from this that the off-diagonal terms in the Einstein tensor vanish rapidly at large times and at small and asymptotically large radius.

Ignoring the off-diagonal components,
the density and (negative) pressure are
\be
    \rho(r,t) = \Lambda  + \frac{3 c_0}{(R a(t))^3} e^{-(r/(R a(t))^2} \quad .
\ee
The Christoffel symbols, ${\Gamma^\mu}_{\alpha \beta}$ vanish at $r=0$ for $\mu$ being either time or radial direction which implies that the metric is spatially flat for small $r$ at fixed positive $t$.  One could also see this from the vanishing of $h(r,t)$ at $r=0$.   This justifies the neglect of the curvature in eq.\ref{eqofcontinuity} near the origin. 
 
The matter density then agrees with eq.\ref{rhom} if we choose 
\be
    c_0 = \frac{M}{3 \pi^{3/2}}
\ee
with $M$ being the total mass of matter in the universe.

 To insure that the energy density reduces to the FRW model in the absence of matter ($c_0=0$), it would be natural to take
\be
     \Lambda = 3 H_v^2
\ee
with the dark energy scale factor 
\be
      a_v(t) = e^{H_v t} \quad .
\ee

In each phase (restoring the dimensional factors of Newton's constant, Planck's constant, and the speed of light), the inflationary expansion is governed by
\be
     {H_v}^2 = \frac{1}{3} \Lambda \frac{8 \pi G_N /c^2}{(\hbar c)^3} \quad .
\ee 
so that
\be
     H_{v1} = 1.44 \cdot 10^{38} {\displaystyle s}^{-1}
\ee
and
\be
     H_{v2} = 1.91 \cdot 10^{-18} {\displaystyle s}^{-1} \quad .
\ee 

\section{A finite mass model with bubble topology}

    In this section we seek a metric that corresponds to matter contained within an expanding bubble.
In this case the metric beyond the bubble boundary is constrained by Birkhoff's theorem to be as in eq.\,\ref{SdSmetric}.  

We retain the form of eq.\,\ref{metric2} for $g_{44}(r,t)$.
However, in order to incorporate a growing bubble of small vacuum energy within a background of high vacuum energy we allow the vacuum energy $\Lambda$ to be discontinuous with the following form:

\be
  \Lambda(r,t) = \Lambda_1 \,\theta(t_2 -t)
 + \theta(t-t_2)\,(\Lambda_1\,\theta(r-t+t_2)
                   + \Lambda_2 \,\theta(t-t_2-r)) \quad .
\label{DarkEnergyTransition}    
\ee
Similarly we would take at positive $t$:
\be
  a(t) = a_1(t) \,\theta(t_2 -t)
 + \theta(t-t_2)\,(a_1(t)\,\theta(r-t+t_2)
                   + a_2(t) \,\theta(t-t_2-r)) 
\label{ScaleFacTransition}    
\ee 
and
\be\nonumber
h(r,t) &=& {(c_m + c_r/a(t))} \frac{r^2}{(R a(t))^3}\,e^{-(r/(R a(t)))^2}\theta(\sqrt{3/2}R a(t)-r)\\
     &+& \theta(r-\sqrt{3/2}R a(t))\, 2 M/r
\quad .
\ee
The photon component proportional to $c_r$ could also be added to the $h$ of the previous section.
 Note that, since $G^4_4$ does not involve time derivatives of the metric and since
we have chosen the scale factor to be continuous, $G^4_4$ does not
acquire delta function singularities except at the bubble surface.

The sharp transitions between phases in eqs.\,\ref{DarkEnergyTransition}$\,$and$\,$\ref{ScaleFacTransition}$\;$should be interpreted as the thin wall limit of
continuous phase transitions described by a hyperbolic tangent function.  
\be
       \theta(x)= \frac{1}{2}\lim_{b\rightarrow 0^+} (1 + \tanh (x/b)) \quad .
\ee
To be consistent one should define $\theta(0)=1/2$.  In the extreme thin wall limit
the deceleration parameter becomes singular at the time of transition suggesting that, in actuality, $b$ should be taken small but non-zero.  Except at the precise
time of transition there is no problem taking $b\rightarrow 0$.

These equations describe a universe beginning at $t=0$ and undergoing a transition from $\Lambda_1$ to a possibly much smaller $\Lambda_2$ at time $t=t_2$.  { Cosmological data on the length of the inflationary era constrains $t_2$.}  Since, in the thin wall limit,  $\Lambda$ is time independent between jumps, the $\theta$ functions modify the previously deduced Einstein tensor only by delta functions at the phase transition jumps and a contribution to the surface tension of the bubble which we ignore.

$M$ is the (finite and constant) total mass inside the bubble. 
Outside the bubble the metric reduces to the Schwarzschild-deSitter form.

Ignoring singular contributions at the bubble boundary the mass density and pressures  are
as before except that the vacuum energy and scale factor have the theta function form of eqs.\,\ref{DarkEnergyTransition} and \ref{ScaleFacTransition}.  Inside the bubble
($r<\sqrt{3/2}\,R\,a(t)$), the densities are
\be
    \rho_v(r,t) = \Lambda(t)
\label{vacdensity}
\ee
\be
    \rho_m(r,t) = \frac{c_m}{(R a(t))^3} e^{-(r/(R a(t))^2}(3-2(r/(a(t)R))^2)
\label{matterdensity}
\ee
\be
    \rho_r(r,t) = \frac{c_r/a(t)}{(R a(t))^3} e^{-(r/(R a(t))^2}(3-2(r/(a(t)R))^2) \quad .
\label{raddensity}
\ee
Integrating numerically, the total mass contained in the bubble is
\be
    M \approx 5.15 c_m 
\label{Mc0relationship}
\ee
differing somewhat from the relation of the previous section.

If, for example, the local matter density were valid up the Hubble length and then dropped to zero, the total mass defining $c_0$ would be 
\be
M \approx \frac{4}{3} \pi L_H^3\,\cdot (1 {\displaystyle{GeV}}/c^2/m^3) \approx  10^{79} {\displaystyle{GeV}}/c^2  \quad .
\label{Mestimate}
\ee

    Observationally, it is known that galaxy cluster counts are vastly lower \cite{Vikhlinin} than predicted from the Planck results.  The observed density of very distant galaxies ($z \approx 7$)
in the Hubble Ultra Deep Field is only a tenth of the local density \cite{Beckwith}.  This could be due to an underestimate of systematic errors since distant galaxies
of low luminosity are undercounted. 
Numerically, the matter density of eq.\,\ref{matterdensity} falls to one tenth of its $r=0$ value at $r=r_0=1.032 R a(t)$.  At present $a(t)=1$.  The relation to redshift is
\be
     r = L_H  \frac{(z+1)^2 -1}{(z+1)^2 + 1}
\ee
with Hubble length $L_H = c a(t)/{\dot{a}(t)}\;$.   $z=7$ corresponds therefore to a distance about $3\%$ less than the Hubble length and thus a value of $R$ about $6\%$ less than the Hubble length could be suggested.  
 If the apparent density drop-off with distance is not due to systematic errors an inhomogeneous model
such as ours would be required.  In the standard cosmological model and in the current model it is an
unexplained coincidence that $H_{v2}$ is approximately the inverse time since the big bang.  In the current model, the closeness of
$R$ to the Hubble length is a similar apparent coincidence.  However, if the current indications of matter inhomogeneity vanish under future re-analysis of systematic errors, our $R$ parameter might have a lower limit much larger than our current fit.  

With the current estimate of the inhomogeneity scale and the observed local matter density (including dark matter) 
\be
 \rho_m(0,t_0) = 1.5\,{\displaystyle GeV/c^2/m^3} = 3 \frac{M}{5.15 R^3}
\ee
we obtain
a fit to $M$, the total mass of the universe:
\be
   M \approx 4.6\,10^{78} {\displaystyle GeV/c^2}
\ee
not far from the estimate of eq.\,\ref{Mestimate}.

If at some future time there is a transition to zero $H_v$ with $p=1$, the boundary of the matter bubble thereafter expands outward with constant speed which can be taken to be the speed of light.

\section{Conclusions}
    In this article we have addressed four conceptual issues in the standard cosmology.  
With respect to the question of infinite cloning, 
we noted previously that this property of the standard model is eliminated if there exists no more than a finite mass.  Similarly, the measure problem of the standard model is avoided since there are no infinite occurences of any event at any given time.
Of course the standard FRW metric with a negative curvature also has a finite total mass and this could be an alternate resolution of the problem.  However the WMAP and Planck results put stringent limits on departures from zero curvature in an FRW analysis.  In the FRW model with spatial curvature, the matter density is constant up to the limiting radius whereas the current model predicts a matter inhomogeneity.  The isotropy of the CBR is often taken to imply matter homogeneity to within one part in $10^5$ but, in fact, observations merely require either that the scale of inhomogeneity, $R$ in the above equations, is above $10^5$ relative to the Hubble length or that the Earth is close to the center of the CBR distribution.  
In ref.\cite{ClavelliGoldstein} we took the first alternative as the default assumption but here we would also like to explore the other possibility.  { This latter possibility may lead to the only way to incorporate a correlation between asymmetries in the CBR and the plane of the solar system \cite{Perivolaropoulos} if the corresponding data survives.}
 
For any finite $R$, the number of identical human clones is finite but, depending on the size of $R$ this number could still be large.  However, it can
be easily checked that, even if there is one human civilization per solar system and $R$ is a billion times the Hubble length, the probability of random cloning of two identical humans is infinitesimal in the current model although it is unity in the standard model.

 It might seem that the present proposal
would involve severe fine tuning and be contrary to the ``Copernican Principle" that the earth does not stand at a privileged position in the universe.  In fact, however, the origin of the Earth's present rest frame is \cite{ClavelliGoldstein} $5.1$ Mpc from the origin of the rest frame of the CBR as determined by the dipole asymmetry in the CBR.  This is about $0.1\%$ of the Hubble length, a minor amount of fine tuning compared to other cosmological coincidences.
In fact, Copernicus did not say there was no center of the universe (solar system) but merely that, at any given time, the earth was displaced from this center by an amount coincidentally similar to $0.1\%$.  
In fact, the main point of Copernicus' discovery is that the motion of multiple planets and a sun becomes mathematically 
simpler in the center of mass frame.  In this sense our proposal is very much in
line with Copernican theory.
On the other hand, the standard cosmology assumes that every observer is slightly displaced from its own center of a spherical shell of background radiation as part of a homogeneous universe.  In this cosmology one would assume that the current rest frame of the earth relative to the CBR is a "peculiar" velocity.

    In the standard model, the universe is expected to be homogeneous at scales above some minimum.  A fractal analysis \cite{Yadov} suggests this minimum is about $370$ Mpc, about that of the Sloan Great Wall and roughly a tenth of the Hubble length.  Any observed structures of greater dimension than this would represent a significant challenge to the standard cosmological model.  Typical clusters of galaxies are $2$ to $3$ Mpc across with large quasar groups (LQG) extending typically up to $200$ Mpc across.  However, recently a huge quasar group  with longest dimension of $1240$ Mpc has been discovered \cite{Clowes} more than three times the expected limit and uncomfortably close to the size of the visible universe.  This calls into question the basic standard assumption of large scale matter homogeneity based on the isotropy of the CBR to within one part in $10^5$.  We feel that this is
another reason to consider inhomogeneous models such as ours although, clearly, our model is not intended to deal with high resolution details such as galactic clusters. 

    With respect to the question of initial conditions, we note that if the matter distribution takes the form of eq.\ref{rhom} with a squared scale factor $a(t)^2$ behaving as a non-integer power of $t$ near the big bang as in eq.\,\ref{scalefac1}, the matter density at negative values of $t$ is complex and everywhere infinite as the time of the big bang is approached from below.  Thus, in this model, time is undefined before the big bang; i.e. time is positive definite.  We suggest that the universe with at least two states of differing vacuum energy is born at the big bang in a state of high vacuum energy with a matter density that is a spatial delta function of position.  Unlike the standard picture in which surviving matter is created out of dark energy at the end of inflation, in our model matter and dark energy are born together in the big bang.  
 For a finite time the matter density dominates over the vacuum energy followed by an inflationary era in which the vacuum energy dominates until a quantum transition to a much lower vacuum energy occurs.
We have discussed elsewhere \cite{ClavelliGoldstein} the chance that there will be a still later transition to a possibly supersymmetric state of zero vacuum energy and constant scale factor in order to avoid the total energy within some radius exceeding the Schwarzschild energy.   

Since the initial state of a physical system is not determined by the equations of motion but must 
merely be one of the states of the system, one cannot seek an explanation for an initial state big bang within physics theory. On the other hand the precise form of 
$a(t)$ might have an explanation within physics and could be different from the simple form we have studied here.  
In this article we have established a consistency between Einstein's equations with a given metric and our previously studied inhomogeneous model with finite matter. 
The model \cite{ClavelliGoldstein} also suggested possible new approaches to other long-standing questions in physics such as the baryon asymmetry problem which are not further discussed here.  Although much work remains to be done along the suggested lines if the impressive though fine-tuned successes of the standard cosmological model are to be reproduced, the current model avoids the
bizarre features and puzzling anomalies of the standard model while fitting observational indications of inhomogeneity.
  
{\bf Acknowledgements}
We thank Tufts cosmologists Larry Ford, Ben Shlaer, Ali Masoumi, and Ken Olum for discussions.

\end{document}

%% file: defs.tex
\usepackage{graphics}
\usepackage[dvips]{color}
\usepackage{multicol}

\usepackage{amsmath,color,fancybox,graphics,graphpap,rotating}
\definecolor{white}{rgb}{1,1,1}
\definecolor{yellow}{rgb}{0.95,0.75,0.1}
\definecolor{red}{rgb}{0.5,0,0}
\definecolor{green}{rgb}{0,1,0}
\definecolor{blue}{rgb}{0,0.5,1}
\definecolor{bgcolor}{rgb}{0.94,0.91,0.78}

\definecolor{lblue}{rgb}{0,0.8,1}
\definecolor{dblue}{rgb}{0,0,.6}
\definecolor{dgreen}{rgb}{0,0.3,0}
\definecolor{lila}{rgb}{0.8,0,0.8}
\definecolor{violet}{rgb}{1,0,1}
\definecolor{grey}{rgb}{0.3,0.3,0.3}
\definecolor{turquoise}{rgb}{0,.9608,1}

\definecolor{contoura}{rgb}{0,0,1}
\definecolor{contourb}{rgb}{0,1,1}
\definecolor{contourc}{rgb}{0,1,0}
\definecolor{contourd}{rgb}{0.95,0.75,0.1}
\definecolor{contoure}{rgb}{1,0,0}
\definecolor{contourf}{rgb}{1,0,1}

\def\lsim{\raise0.3ex\hbox{$\;<$\kern-0.75em\raise-1.1ex\hbox{$\sim\;$}}}
\def\gsim{\raise0.3ex\hbox{$\;>$\kern-0.75em\raise-1.1ex\hbox{$\sim\;$}}}

\newcommand{\nee}{\nonumber \end{eqnarray}}
